\documentclass[10pt,a4paper]{article}
\usepackage{amsmath}
\usepackage{amssymb}
\usepackage{amsfonts}
\usepackage{amstext}
\usepackage{amsbsy}
\usepackage[mathscr]{eucal}
\usepackage{graphicx}
\newcommand{\beq}{\begin{equation}}
\newcommand{\eeq}{\end{equation}}
\newcommand{\beqa}{\begin{eqnarray}}
\newcommand{\eeqa}{\end{eqnarray}}
\newcommand{\ket}[1]{| #1 \rangle}

\makeindex
\title{\Large\textbf{Schwarz inequality and concurrence}}

\author{\textit{Hoshang Heydari$^{\dagger}$}  and \textit{Gunnar Bj\"{o}rk}\\
       \small \textit{hoshang@imit.kth.se, http://www.imit.kth.se/QEO/}\\
        \small\textit{Department of Microelectronics and}
 \small\textit{Information Technology,}\\
 \small\textit{Royal Institute of Technology (KTH)},
\small\textit{Electrum 229, SE-164 40 Kista, Sweden}\\
\small\textit{$^{\dagger}$Nihon University, Institute of Quantum
Science, Quantum Optics Group,}\\ \small\textit{1-8,
Kanda-Surugadai, Chiyoda-ku, Tokyo 101- 8308, Japan }}
\date{}
\begin{document}

\maketitle \thispagestyle{empty}

\maketitle
\begin{abstract}
We establish a relation between the Schwarz inequality and the
generalized concurrence of an arbitrary, pure, bipartite or
tripartite state. This relation places concurrence in a
geometrical and functional-analytical setting.
\end{abstract}

\section{Introduction}
Quantum entanglement is one of the most interesting and debated
properties of quantum mechanics. It has become an essential resource
for the quantum communication created in recent years, with some
potential applications such as quantum cryptography
\cite{Bennett84,Ekert91} and quantum teleportation \cite{Bennett93}.
The idea of quantum entanglement goes back to the early days of
quantum theory where it was initiated by Schr\"{o}dinger, Einstein,
Podolsky and Rosen \cite{Sch35,EPR35} and was later extended by Bell
\cite{Bell64} in the form of Bell inequalities. Quantification of
multipartite state entanglement \cite{Lewen00,Dur99} is difficult
and is a task that is directly linked to  linear algebra, geometry
and functional analysis. The definition of separability and
entanglement of a multipartite state was introduced in
\cite{Vedral97} following the definition for bipartite states, given
in 1989 by Werner \cite{Werner89}. One widely used measure of
entanglement for a pair of qubits, is entanglement of formation
\cite{Bennett96}. A closely related measure is concurrence, that
gives an analytic formula for the entanglement of formation
\cite{Wootters98}. In recent years, there have been several
proposals to generalize this measure to general bipartite states,
e.g., Uhlmann \cite{Uhlmann00} has generalized the concept of
concurrence by considering arbitrary conjugation, then Audenaert,
Verstraete, and De Moor \cite{Audenaert} generalized this formula in
spirit of Uhlmann's work, by defining a concurrence vector for pure
states. Another generalization of concurrence have been done by
Rungta \emph{et al.} \cite{Rungta01} based on an idea of a
superoperator called universal state inversion. And finally Gerjuoy,
Albeverio and Fei, Akhtarshenas, and Bhaktavatsala and Ravishankar
\cite{Gerjuoy,Albeverio,Akhtarshenas,Bhaktavatsala} have given
explicit expression in terms of the state amplitude coefficient of a
pure bipartite state in any dimension.

In this paper, we put the concurrence in another perspective, namely
we establish a relation between Schwarz' inequality and concurrence
for bipartite states and then extend our connection to multipartite
states. We show that, the generalized concurrence \cite{Albeverio}
and entanglement tensor \cite{Hosh4} for a three-partite state can
be derived using the concept of Schwarz inequality. Generalization
of this relation to a multipartite state with more than three
subsystems can be tried out in the same way as for three-partite
state but it gives only information about the set of separable state
and can not quantify a general pure multipartite state completely.

%%%%%%%%%%%%%%%%%%%%%%%%%%%%%%%%%%%%%%%%%%%%%%%%%%%%%%%%%%%%%%%%
\section{Entanglement}
In this section we will establish the notation for separable states
and entangled states. Let us denote a general, pure, composite
quantum system with $m$ subsystems
$\mathcal{Q}=\mathcal{Q}^{p}_{m}(N_{1},N_{2},\ldots,N_{m})
=\mathcal{Q}_{1}\mathcal{Q}_{2}\cdots\mathcal{Q}_{m}$, consisting of
a state
\begin{equation}\label{Mstate}
\ket{\Psi}=\sum^{N_{1}}_{i_{1}=1}\sum^{N_{2}}_{i_{2}=1}\cdots\sum^{N_{m}}_{i_{m}=1}
\alpha_{i_{1},i_{2},\ldots,i_{m}} \ket{i_{1},i_{2},\ldots,i_{m}}
\end{equation}
 defined on a Hilbert space
\begin{eqnarray}
\mathcal{H}_{\mathcal{Q}}&=&\mathcal{H}_{\mathcal{Q}_{1}}\otimes
\mathcal{H}_{\mathcal{Q}_{2}}\otimes\cdots\otimes\mathcal{H}_{\mathcal{Q}_{m}}\\\nonumber
&=&\mathbf{C}^{N_{1}}\otimes\mathbf{C}^{N_{2}}\otimes\cdots\otimes\mathbf{C}^{N_{m}},
\end{eqnarray}
where the dimension of the $j$th Hilbert space is given  by
$N_{j}=\dim(\mathcal{H}_{\mathcal{Q}_{j}})$. We are going to use
this notation throughout this paper, i.e., we denote a pure pair of
qubits by $\mathcal{Q}^{p}_{2}(2,2)$. Next, let $\rho_{\mathcal{Q}}$
denote a density operator acting on $\mathcal{H}_{\mathcal{Q}}$. The
density operator $\rho_{\mathcal{Q}}$ is said to be fully separable,
which we will denote by $\rho^{sep}_{\mathcal{Q}}$, with respect to
the Hilbert space decomposition, if it can  be written as
\begin{equation}\label{eq:sep}
\rho^{sep}_{\mathcal{Q}}=\sum^\mathrm{K}_{k=1}p_k
\bigotimes^m_{j=1}\rho^k_{\mathcal{Q}_{j}},~\sum^N_{k=1}p_{k}=1
\end{equation}
 for some positive integer $\mathrm{K}$, where $p_{k}$ are positive real
numbers and $\rho^k_{\mathcal{Q}_{j}}$ denote a density operator on
Hilbert space $\mathcal{H}_{\mathcal{Q}_{j}}$. If
$\rho^{p}_{\mathcal{Q}}$ represents a pure, fully separable state,
then $\mathrm{K}=1$. If a state is not fully separable, then it is
called an entangled state. A completely nonseparable quantum system
is one that in any basis must be written
\begin{equation}\label{eq:completely}
\rho^{nonsep}_{\mathcal{Q}}=\sum^\mathrm{K}_{k=1}p_k
\rho^k_{\mathcal{Q}},~\sum^N_{k=1}p_{k}=1,
\end{equation} where $\mathcal{Q}=\mathcal{Q}^p_1(N_1 N_2 \cdots
N_m)$. The simplest such completely nonseparable, generic states
(that, moreover, are maximally entangled) are Bell states and
$\mathrm{W}$-states.

%%%%%%%%%%%%%%%%%%%%%%%%%%%%%%%%%%%%%%%%%%%%%%%%%%%%%%%%%%%%%%%%%%%%%%%%%%%%%%%%%%%%%%%%%%%%%%%%
\section{The Schwarz inequality and concurrence} \label{Analytical}

In this section, we will investigate the relation between
concurrence, the Schwarz inequality, and the minors of determinant
of bipartite states, which are directly related to the geometry of
the Hilbert space and Segre variety \cite{Hosh5}.

Let us begin by reviewing the Schwarz inequality  on an inner
product space such as a Hilbert space, and then use this inequality
to relate it to the geometry of concurrence.

Let $X_{1}=(\xi_{1},\xi_{2})$ and $X_{2}=(\eta_{1},\eta_{2})$ be two
vectors defined on the complex Hilbert space
$\mathcal{H}=\mathbf{C}^{2}$. Then $X_{1}$ and $X_{2}$ are parallel
if and only if $\left|\xi_{1}\eta_{2}-\eta_{1}\xi_{2}\right|=0$. We
will prove this using the Schwarz inequality $\langle
X_{1}\ket{X_{2}}\langle
X_{2}\ket{X_{1}}\leq\|X_{1}\|^{2}\cdot\|X_{2}\|^{2}$ as follows:
\begin{eqnarray}
\langle X_{1}\ket{X_{2}}\langle
X_{2}\ket{X_{1}}&=&(\xi_{1}\bar{\eta}_{1}+\xi_{2}\bar{\eta}_{2})
(\bar{\xi}_{1}\eta_{1}+\bar{\xi}_{2}\eta_{2})\\\nonumber &=&
|\xi_{1}|^{2} |\eta_{1}|^{2}
+\xi_{1}\bar{\eta}_{1}\bar{\xi}_{2}\eta_{2}
+\xi_{2}\bar{\eta}_{2}\bar{\xi}_{1}\eta_{1}+|\xi_{2}|^{2}
|\eta|_{2}^{2},
\end{eqnarray}
where, i.e., $\bar{\xi}$ is the complex conjugate of $\xi$. The
product of the norms of these vectors is given by
\begin{equation}
%%%%%%%%%%%%%%%%%%
\|X_{1}\|^{2}\cdot\|X_{2}\|^{2} =
 |\xi_{1}|^{2} |\eta_{1}|^{2}+|\xi_{1}|^{2} |\eta_{2}|^{2}+ |\xi_{2}|^{2} |\eta_{1}|^{2}+|\xi_{2}|^{2}
 |\eta_{2}|^{2}.
\end{equation}
If $X$ and $Y$ are  parallel, then we have $ \langle
X_{1}\ket{X_{2}}\langle
X_{2}\ket{X_{1}}=\|X_{1}\|^{2}\cdot\|X_{2}\|^{2} $, which implies
that
\begin{eqnarray}
\xi_{1}\bar{\eta}_{1}\bar{\xi}_{2}\eta_{2}
+\xi_{2}\bar{\eta}_{2}\bar{\xi}_{1}\eta_{1} &=&
%%%
|\xi_{1}|^{2} |\eta_{2}|^{2}+ |\xi_{2}|^{2} |\eta_{1}|^{2}
\Longrightarrow \left| \xi_{1}\eta_{2}-\eta_{1}\xi_{2} \right|^{2} =
0.
\end{eqnarray}
That is, $X_{1}$ and $X_{2}$ are  parallel if, and only if,
\beq \det\left(%
\begin{array}{cc}
  \xi_{1} &\xi_{2} \\
  \eta_{1} &  \eta_{2}\\
\end{array}%
\right)=0 , \eeq where $\det$ denotes the determinant. We note that
the area of a parallelogram spanned by two vectors is equal to the
value of their 2-by-2 determinant.

Now we set out to generalize this simple result to a larger
bipartite product space. Let
$X_{1}=(\xi_{1},\xi_{2},\ldots,\xi_{N_{2}})$ and
$X_{2}=(\eta_{1},\eta_{2},\ldots,\eta_{N_{2}})$ be two vectors
defined on the  Hilbert space $\mathcal{H}=\mathbf{C}^{N_{2}}$.
Again by using the Schwarz inequality, we get
\begin{eqnarray}
\langle X_{1}\ket{X_{2}}\langle
X_{2}\ket{X_{1}}&=&\nonumber(\xi_{1}\bar{\eta}_{1}+\xi_{2}\bar{\eta}_{2}+\ldots+\xi_{N_{2}}\bar{\eta}_{N_{2}})
(\bar{\xi}_{1}\eta_{1}+\bar{\xi}_{2}\eta_{2}+\ldots+\bar{\xi}_{N_{2}}\eta_{N_{2}})\\\nonumber
&=& |\xi_{1}|^{2} |\eta_{1}|^{2}
+\xi_{1}\bar{\eta}_{1}\bar{\xi}_{2}\eta_{2}
+\xi_{1}\bar{\eta}_{1}\bar{\xi}_{3}\eta_{3}
+\xi_{2}\bar{\eta}_{2}\bar{\xi}_{1}\eta_{1} +|\xi_{2}|^{2}
|\eta_{2}|^{2}\\\nonumber && +\ldots+
 \xi_{N_{2}-1}\bar{\eta}_{N_{2}-1}\bar{\xi}_{N_{2}}\eta_{N_{2}}
+\xi_{N_{2}}\bar{\eta}_{N_{2}}\bar{\xi}_{N_{2}-1}\eta_{N_{2}-1}
\\
&&+|\xi_{N_{2}}|^{2}|\eta_{N_{2}}|^{2},
\end{eqnarray}
and, in the same way as we have done above, we calculate the product
of the norms of these vector as follows:
\begin{eqnarray}
 \|X_{1}\|^{2}\cdot\|X_{2}\|^{2}&=&\nonumber
 |\xi_{1}|^{2}(|\eta_{1}|^{2}+|\eta_{2}|^{2}+\ldots+|\eta|_{N_{2}}^{2})\\\nonumber
&&+
|\xi_{2}|^{2}(|\eta_{1}|^{2}+|\eta_{2}|^{2}+\ldots+|\eta_{N_{2}}|^{2})
 \\\nonumber
&&+
\ldots+|\xi_{N_{2}}|^{2}(|\eta_{1}|^{2}+|\eta_{2}|^{2}+\ldots+|\eta_{N_{2}}|^{2})
 \\\nonumber
&=&
 |\xi_{1}|^{2}|\eta_{1}|^{2}+|\xi_{1}|^{2}|\eta_{2}|^{2}+\ldots+|\xi_{1}|^{2}|\eta_{N_{2}}|^{2}
 \\\nonumber
&&+
|\xi_{2}|^{2}|\eta_{1}|^{2}+|\xi_{2}|^{2}|\eta_{2}|^{2}+\ldots+|\xi_{2}|^{2}|\eta_{N_{2}}|^{2}
\\&&
 +
 \ldots+|\xi_{N_{2}}|^{2}|\eta_{1}|^{2}+|\xi_{3}|^{2}|\eta_{2}|^{2}+\ldots+|\xi_{N_{2}}|^{2}|\eta_{N_{2}}|^{2}.
\end{eqnarray}
Again, if $X_{1}$ and $X_{2}$ are parallel, then we have $\langle
X_{1}\ket{X_{2}}\langle
X_{2}\ket{X_{1}}=\|X_{1}\|^{2}\cdot\|X_{2}\|^{2}$ which, after some
simplification, can be rewritten as follows:
\begin{eqnarray}
&&\nonumber
|\xi_{1}|^{2}|\eta_{2}|^{2}-\xi_{1}\eta_{2}\bar{\eta}_{1}\bar{\xi}_{2}
-\xi_{2}\eta_{1}\bar{\eta}_{2}\bar{\xi}_{1}+|\xi_{2}|^{2}|\eta_{1}|^{2}
+\ldots+|\xi_{N_{2}-1}|^{2}|\eta_{N_{2}}|^{2}\\\nonumber&&-\xi_{N_{2}-1}\eta_{N_{2}}\bar{\eta}_{N_{2}-1}\bar{\xi}_{N_{2}}
-\xi_{N_{2}}\eta_{N_{2}-1}\bar{\eta}_{N_{2}}\bar{\xi}_{N_{2}-1}+|\xi_{N_{2}}|^{2}|\eta_{N_{2}-1}|^{2}\\\nonumber
&&=\left|\xi^{2}_{1}\eta^{2}_{2}-\xi_{1}\eta_{2}\right|^{2}
+\ldots+\left|\xi_{N_{2}-1}\eta_{N_{2}}
-\xi_{N_{2}}\eta_{N_{2}-1}\right|^{2}=0.
\end{eqnarray}

That is, $X_{1}$ and $X_{2}$ are parallel if, and only if,
\begin{equation}
\left|\xi_{1}\eta_{2}-\eta_{1}\xi_{2}\right|
=\left|\xi_{1}\eta_{3}-\eta_{1}\xi_{3}\right|
=\cdots=\left|\xi_{N_{2}-1}\eta_{N_{2}}-\eta_{N_{2}-1}\xi_{N_{2}}\right|=0.
\end{equation}
This result implies that
$$
\det\left(%
\begin{array}{cc}
  \xi_{1} &\xi_{2} \\
  \eta_{1} &  \eta_{2}\\
\end{array}%
\right) =\cdots
=\det\left(%
\begin{array}{cc}
  \xi_{N_{2}-1} &\xi_{N_{2}} \\
  \eta_{N_{2}-1} &  \eta_{N_{2}}\\
\end{array}%
\right)=0 $$ if the vectors are parallel.
%%%%%%%%%%%%%%%%%%%%%%%%%%%

To establish a relation between the Schwarz inequality and the
concurrence, let us consider the quantum system
$\mathcal{Q}^{p}_{2}(N_{1},N_{2})$ be a pure, bipartite quantum
system. Then, the concurrence can be written as \cite{Albeverio}
\beq \mathcal{C}(\mathcal{Q}^{p}_{2}(N_{1},N_{2})) =
\left(\mathcal{N}\sum_{l_1>k_1}^{N_1}\sum_{k_1=1}^{N_1}\sum_{
l_2>k_2}^{N_2}\sum_{k_2=1}^{N_2}|\mathrm{T} \left (
\begin{array}{cc} k_1 & l_1 \\ k_2 & l_2
\end{array} \right )|^{2}\right)^{\frac{1}{2}}
\label{eq:concurrence} \eeq where $\mathrm{T} \left (
\begin{array}{cc} k_1 & l_1 \\ k_2 & l_2
\end{array} \right ) = \det\left ( \begin{array}{cc} \alpha_{k_1, k_2} &  \alpha_{k_1, l_2}\\
 \alpha_{l_1, k_2} & \alpha_{l_1, l_2}
\end{array} \right ) $ is a second order minor of the $2 \times N_{2}$ matrix
%%%%%%%%%%%%%%%%%%%%%%%%%%%%%%%%%%%%%%%%%
\begin{equation}\label{eq:pq}
\left(%
\begin{array}{ccccc}
  \alpha_{k_1,1} &\alpha_{k_1,2} & \cdots & \alpha_{k_1,N_{2}-1} &\alpha_{k_1,N_{2}}\\
  \alpha_{l_1,1} &\alpha_{l_1,2} & \cdots & \alpha_{l_1,N_{2}-1} &\alpha_{l_1,N_{2}}\\
\end{array}%
\right),
\end{equation}
%%%%%%%%%%%%%%%%
%%%%%%%%%%%%%%%%%%%%%%%
where $\mathcal{N}$ is a normalization constant.
%%%%%%%%%%%%%%%%%%%%%%%%%%%%%%%%%%%%%%%%%%%%%%%
We recognize the expression (\ref{eq:concurrence}) as the sum of
all parallelograms computed above. Hence, the concurrence is zero
only if the Schwartz inequality is satisfied with equality for all
pairs of vectors
$X_{k_1}=(\alpha_{k_1,1},\alpha_{k_1,2},\ldots,\alpha_{k_1,N_{2}})$
and
$X_{l_1}=(\alpha_{l_1,1},\alpha_{l_1,2},\ldots,\alpha_{l_1,N_{2}})$.
This implies that all the vectors $X_{k_1}$ and $X_{l_1}$ are
parallel. If so, the state is obviously separable because this
means that the state can be written \beq
\ket{\Psi}=(\alpha_{1,1}\ket{1_1} + \ldots +
\alpha_{N_1,1}\ket{{N_1}_1}) \otimes (\ket{1_2}+
\alpha_{1,2}/\alpha_{1,1} \ket{2_2} + \ldots +
\alpha_{1,N_{2}}/\alpha_{1,1} \ket{{N_2}_2}). \eeq We also see
that the concurrence for a bipartite, pure state, loosely
speaking, has the geometrical interpretation of the summed
pairwise deviation from parallelism of all the vectors $X_{k_1}$
and $X_{l_1}$.

\section{The Schwarz inequality and concurrence of a general pure
three-partite state}
%%%%%%%%%%%%%%%%%%%%%%%%%%%%%%%%
Let us now see what happens if we consider the simplest example of
a three-partite system.  The simplest tripartite system,
consisting of three qubits, is denoted
$\mathcal{Q}^{p}_{3}(2,2,2)$. The concurrence of this state is
then given by \cite{Albeverio} \beq
\mathcal{C}(\mathcal{Q}^{p}_{3}(2,2,2)) =  \left( \mathcal{N}
\sum^{3}_{j=1}\sum_{ l>k}^{2,2}\sum_{k=1,1}^{2,2}\sum_{l_j>k_j}^2
\sum_{k_j=1}^{2} |\mathrm{T} \left (
\begin{array}{cc} k_j & l_j \\ k \neq k_j & l \neq l_j
\end{array} \right )|^{2}\right)^{\frac{1}{2}}\eeq
%%%%%%%%%%%%%%%%%
Where $\mathrm{T}\left (
\begin{array}{cc} k_1 & l_1 \\ k \neq k_1 & l \neq l_1
\end{array} \right )$ is a minor of the $2 \times 4$ matrix
\begin{equation}
\left(%
\begin{array}{cccc}
  \alpha_{k_1,1,1} & \alpha_{k_1,1,2}&\alpha_{k_1,2,1}&\alpha_{k_1,2,2} \\
 \alpha_{l_1,1,1} & \alpha_{l_1,1,2}&\alpha_{l_1,2,1}&\alpha_{l_1,2,2} \\
\end{array}%
\right),
\end{equation} and where the two-digit indices $k$ and $l$ run
from $1,1$ to $2,2$, and, where of course the only possibility is
that $k_1=1$ and $l_1=2$. In the same manner, $\mathrm{T}\left (
\begin{array}{cc} k_3 & l_3 \\ k \neq k_3 & l \neq l_3
\end{array} \right )$ is a minor of the matrix
\begin{equation}
\left(%
\begin{array}{cccc}
  \alpha_{1,1,k_3} & \alpha_{1,2,k_3}&\alpha_{2,1,k_3}&\alpha_{2,2,k_3} \\
 \alpha_{1,1,l_3} & \alpha_{1,2,l_3}&\alpha_{2,1,l_3}&\alpha_{2,2,l_3} \\
\end{array}%
\right) ,
\end{equation}
etc.

If we apply the Schwarz inequality to all the combinations of the
above pairs of vectors, then we get the desired result. The
interpretation of the result is that $\mathrm{T} \left (
\begin{array}{cc} k_j & l_j \\ k \neq k_j & l \neq l_j
\end{array} \right )$ generates the
minor determinants that establish whether system
$\mathcal{Q}^{p}_j$ is separable from the rest of the system.
%%%%%%%%%%%%%%%%%%%%%%%%%%%%%%%%%%%
Again we can easily generalize the result above to a general,
pure, three-partite state
$\mathcal{Q}^{p}_{3}(N_{1},N_{2},N_{3})$. Then concurrence of this
state is given by \cite{Albeverio} \beq
\mathcal{C}(\mathcal{Q}^{p}_{3}(N_{1},N_{2},N_{3}))=\left(
\mathcal{N} \sum^{3}_{j=1}\sum_{ l>k}\sum_{k}\sum_{
l_j>k_j}^{N_j}\sum_{k_j=1}^{N_j}|\mathrm{T}\left (
\begin{array}{cc} k_j & l_j \\ k \neq k_j & l \neq l_j
\end{array} \right )|^{2}\right)^{\frac{1}{2}}\eeq
where, e.g., the indices $k$ and $l$ for $j=1$ run through the
$N_2 N_3$, two-digit numbers $1,1$ to $N_2,N_3$.
%%%%%%%%%%%%%%%%%%%%%%%%%

From the above discussion we can see that the equality in Schwarz
inequality can be used as a criterion for separability, and the
deviation from equality, in the sense outlined above, can be used as
measure of entanglement which coincides with generalized concurrence
and our entanglement tensor for bi- and three-partite states.
%%%%%%%%%%%%%%%%%%%%%%%%
\section{Conclusion}

We  have discussed the relation between Schwarz inequality (or,
rather, equality) and concurrence for bi- and three-partite states
and possible generalization to multi-partite states. This relation
helps us visualize the geometrical properties of concurrence and the
relation is directly related to the geometry of the Hilbert space as
a normed complex space with an inner-product defined on it.
Moreover, we have shown that the deviation from the Schwarz
inequality upper bound (perhaps this bound can be called the
``Schwarz equality'') can be used as measure of entanglement for
concurrence for bi- and three-partite states.

\begin{flushleft}
\textbf{Acknowledgments:} This work was supported by the Swedish
Foundation for Strategic Research, the Swedish Research Council, and
the Wenner-Gren Fundations.
\end{flushleft}

%%%%%%%%%%%%%%%%%%%%%%%%%%%%%%%%%%%
%%%%%%%%%%%%%%%%%%%%%%%%%%%%%%%%%%%%%%%%%%%%%%%%%%%%%%%%%%%%%%%%%%%%%%%%%%%%%%%%%%%%%

%%%%%%%%%%%%%%%%%%%%%%%%%%%%%%%

\end{document}